\documentclass[10pt,a4paper]{article}

\usepackage[T1]{fontenc} 

\usepackage[english]{babel} 

\usepackage{abstract} 

\usepackage{titling} 

\usepackage{hyperref} 
\usepackage[square,numbers]{natbib}

\usepackage{graphicx}

\title{\LARGE \bf Response of solar irradiance \\ to sunspot area variations}

\author{Thierry Dudok de Wit$^1$, Greg  Kopp$^{2,3}$, Alexander Shapiro$^3$, Veronika Witzke$^3$, \\
Matthieu Kretzschmar$^1$ \\[1ex] 
\small $^1$ LPC2E, University of Orl\'eans and CNRS, Orl\'eans, France \\ 
\small $^2$ Laboratory for Atmospheric and Space Physics, University of Colorado, Boulder, CO, USA \\ 
\small $^3$ Max-Planck-Institut f\"ur Sonnensystemforschung, G\"ottingen, Germany \\
}

\renewcommand{\maketitlehookd}{%
\begin{abstract}
\noindent One of the important open questions in solar irradiance studies is whether long-term variability (i.e. on timescales of years and beyond) can be reconstructed by means of models that describe short-term variability (i.e. days) using solar proxies as inputs. Preminger and Walton (2005, GRL, 32, 14109) showed that the relationship between spectral solar irradiance and proxies of magnetic-flux emergence, such as the daily sunspot area, can be described in the framework of linear system theory by means of the impulse response.  We significantly refine that empirical model by removing spurious solar-rotational effects and by including an additional term that captures long-term variations. Our results show that long-term variability cannot be reconstructed from the short-term response of the spectral irradiance, which cautions the extension of solar proxy models to these timescales. In addition, we find that the solar response is nonlinear in such a way that cannot be corrected simply by applying a rescaling to sunspot area.

\end{abstract}
}

\date{\normalsize This is an updated version of the article originally published as:  Dudok de Wit, T., Kopp, G., Shapiro, A., Witzke, V. and Kretzschmar, M. \textit{Response of Solar Irradiance to Sunspot-area Variations}, The Astrophysical Journal (2018) 853:197, \ \ 
\url{http://dx.doi.org/10.3847/1538-4357/aa9f19}} 

\begin{document}

\maketitle

\section{Introduction}

The variability of solar irradiance was unambiguously proven almost four decades ago, after the launch of the NIMBUS~7 mission \citep{hickey80}. Even prior to this, it was suspected that a major contributor to climate change could be caused by a  variable Sun \citep{eddy76}, after which studies of solar irradiance variability attracted considerable attention.
Over the last few years significant progress has been achieved in understanding the physical causes of solar variability and the mechanisms by which it can influence Earth's climate. The mechanisms causing such irradiance variations have largely been identified \citep{domingo09,solanki13,yeo17} with magnetic-field evolution, and solar rotation causing variations on timescales larger than a day while granulation and oscillations are responsible for variations on shorter timescales. 

An uninterrupted series of space-based missions beginning with the NIMBUS~7 have monitored the variability of the total solar irradiance (TSI), which is the spectrally-integrated solar-radiative flux normalized to one AU from the Sun, with high precision and accuracy \citep{kopp16,ddw17}. Current TSI models can  reproduce these measurements well on timescales from minutes to decades  \citep{seleznyov11,yeo14,coddington16,shapiro17}. Daily measurements of the spectral solar irradiance (SSI) in the ultraviolet began at about the same time and have since been extended to span the visible and near-infrared up to 2.4 $\mu m$.

However, a number of important open questions remains. In particular, while models and observations of the variability of the SSI agree reasonably well short-term (i.e.~on timescales from a few days to solar-rotational periods) \citep{ermolli13,solanki13}, the amplitude and even the phase of the variations on timescales of the 11-year solar-activity cycle and longer are still unclear \citep{harder09,wehrli11}. Knowledge of such long-term SSI variations is crucial for assessing the impact of solar variability on the Earth's climate. The lack of observational constraints on their magnitude has unfortunately hampered Sun-climate studies \citep[e.g.][]{oberlander12,ball16}.

In order to solve this problem, numerous attempts have been made to extrapolate SSI variations to multi-decadal timescales by means of empirical models that reproduce the dynamics at shorter timescales \citep{lean00b,preminger06b,coddington16}. Different physical mechanisms may affect the SSI variability on different timescales, however \citep{shapiro15}, making such extrapolations non-trivial. Nevertheless, such irradiance models appear to reproduce the observations reasonably well short-term, at least over the four decades for which direct observations are available. Their ability to properly reproduce longer-term variations is an open and hotly-debated question that remains unanswered partly because of a lack of stable, long-term SSI measurements, particularly in the visible and near-infrared spectral bands where relative solar-variability is small.

Irradiance-reconstruction models rely on several assumptions. First, they assume that the SSI can be reconstructed from a linear or weakly nonlinear combination of one or more solar proxies such as sunspot number or sunspot area. A second assumption is that the relation between the SSI and proxies is time-invariant, which is needed in order to reconstruct the SSI backward in time before the observations began. However, the models are determined using data from only the relatively-short overlap period (typically one to four decades) during which both the SSI observations and proxies are available, and thus rely on time-invariance for extrapolations over their multi-century pre-observation time ranges. A third assumption, which is an important focus of this article, is the instantaneity of the response of the SSI to solar proxies, particularly to flux emergence as indicated by changes in the sunspot area. Preminger et al. \citep{preminger05,preminger06a} were among the first to consider the convolutive nature of this particular response and the possibility of a time delay relative to changes in the proxies to which they are causally related. Incorporating such a delay improves the impulse-response models' reconstruction capacity and also gives deeper insight into their physical interpretation. 

The usual framework for describing such a non-instan\-tane\-ous response between an input (a solar proxy) and an output (the SSI) is that of linear systems with transfer functions that are classically represented by their impulse response \citep{antsaklis05}. The impulse response relating the SSI to a solar proxy such as the sunspot area indicates by how much an increase in sunspot area leads to an excess or deficit of spectral irradiance. Such relations may help determine whether solar brightness variations are spot- or faculae-dominated, which in turn helps understand solar variability on longer timescales. Several recent studies have been devoted to this question \citep{preminger11,woods15,shapiro16}.

Motivated by this important question, we evaluate here the impulse response of the SSI to the sunspot area along the line initiated by  \citep{preminger05,preminger06a}. However, we carry the study further by 1) eliminating solar-rotational effects that hamper the proper estimation of the impulse response and its connection to the energy budget, and 2) introducing a more general transfer function model that allows separating causally-related effects from other contributions. Our results reveal the nonlinearity of the impulse response during the phase when there is flux emergence (i.e. before the sunspot decays) and reveal that it not possible to reconstruct long-term variability from a linear impulse-response model based on short-term measurements. We stress that our objective is to infer physical information from the impulse response and not to find the best model for reconstructing the SSI from sunspots, which is a different problem.

The paper is organized as follows: the methodology is described in Section~\ref{sec_methodology}, followed by the description of the data sets in Section~\ref{sec_data}. We discuss the results in Section~\ref{sec_discussion} with the conclusions summarized in Section~\ref{sec_conclusions}.



\section{Methodology}
\label{sec_methodology}

\subsection{Impulse Response with no Residual}

Our approach is conceptually similar to that introduced by Preminger et al. \citep{preminger05,preminger06a,preminger06b,preminger07} (henceforth denoted as PW), with two major differences that are described in Sections~\ref{subsect:res} and \ref{subsect:rot}. The studies by PW suggested the possibility of describing the relationship between variations in the sunspot area and in the SSI by means of a linear time-invariant system. An essential feature of such time-invariant systems is that they are entirely characterized by their impulse response $h(t)$, which relates their response $y(t)$ to a given stimulus or input $u(t)$ by
\begin{equation}
y(t) = h(t)*u(t) = \int_{-\infty}^{\infty} h(t-\tau) \; u(\tau) \; \mathrm{d}\tau \ .
\label{eq:impulse1}
\end{equation}
The Fourier transform turns this convolution into a product
\begin{equation}
Y(\omega) = H(\omega) U(\omega),
\label{eq:impulse2}
\end{equation}
where $Y(\omega)$ stands for the Fourier transform of $y(t)$, etc. Once we have estimated the impulse response $h(t)$ or the transfer function $H(\omega)$, we can model the response to any type of temporal evolution of the input. The impulse response may be estimated by working in Fourier space: for each frequency $\omega$ we then perform a linear regression of $Y(\omega)$ versus $U(\omega)$.  

\begin{figure}[!htb] 
    \begin{center}
    \includegraphics[width=0.55\textwidth]{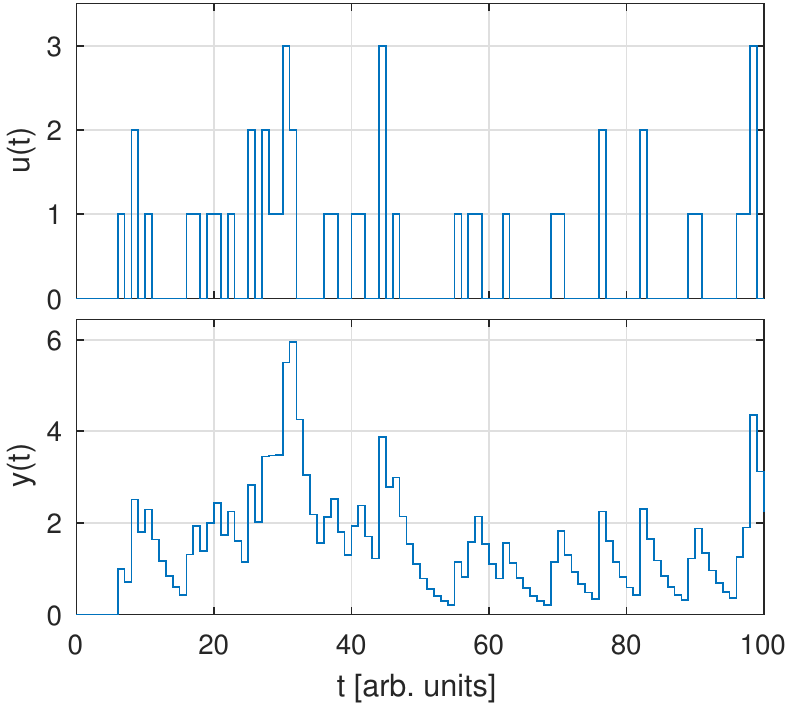}
    \end{center}
    
    \caption{Example showing an input $u(t)$ that consists of random pulses, and the associated response $y(t)$ for the model described by equation~\ref{eq:dynsyst} with a relaxation time $\tau = 4$. \label{fig_simul1}}
\end{figure}

To illustrate this concept let us imagine a simple dynamical system that is described by the linear differential equation
\begin{equation}
\frac{\textrm{d}}{\textrm{d}t} y(t) = u(t) - \frac{y(t)}{\tau}.
\label{eq:dynsyst}
\end{equation}
For a discrete time series this first-order differential equation can be reproduced by a simple first-order autoregressive model with exogenous inputs \citep{ljung97}. As an example, Figure~\ref{fig_simul1} shows an input $u(t)$ that consists of a series of discrete random pulses; these could be, for instance, a time series of the sunspot number near solar minimum. The shown modeled output $y(t)$ computed from equation~\ref{eq:dynsyst} 
tracks the input with a relaxation decay time $\tau$. This output is qualitatively similar to what we shall observe later for that of the UV irradiance to changes in sunspot area.

\begin{figure}[!htb] 
    \begin{center}
    \includegraphics[width=0.55\textwidth]{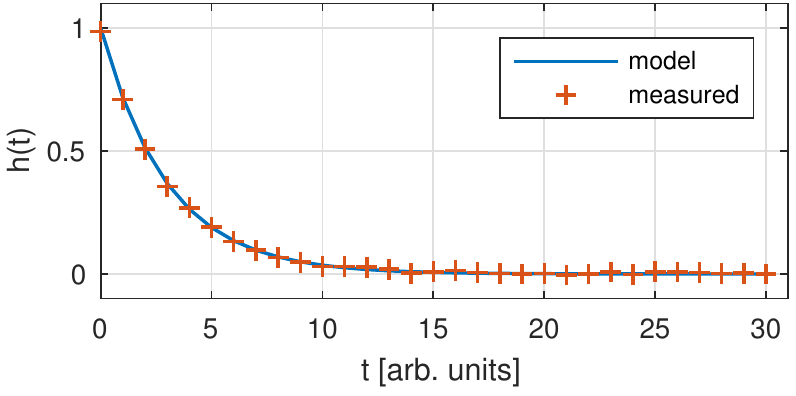}
    \end{center}
    
    \caption{Illustration of the excellent agreement between the theoretical impulse response $h(t)$ for the model described by equation~\ref{eq:dynsyst} (blue line) and that obtained by computing the inverse Fourier transform of $H(\omega) = Y(\omega) / U(\omega)$ (red plus signs) using the data from the example shown in Figure~\ref{fig_simul1} after adding noise to them. \label{fig_simul2}}
\end{figure}

Taking the Fourier transforms of the in- and out-put from the example in Figure~\ref{fig_simul1}, we estimate the impulse response $h(t)$, which is illustrated in Figure~\ref{fig_simul2} together with its theoretical model based on equation~\ref{eq:dynsyst}. 

We simulated the dynamical system in this example using conditions that are similar to the ones encountered with real solar data, such as the length of the data sequences and an inclusion of \textasciitilde20\% additive white noise in the in- and out-put. As shown in Figure~\ref{fig_simul2}, the reconstructed impulse response agrees remarkably well with the theoretical one, in spite of the contamination of the data by noise. The  estimation of such impulse responses from finite and noisy records is a central task in the field of system identification \citep{ljung97} and a considerable amount of literature has been devoted to the design of robust estimators, with many applications in automatic control, signal processing, telecommunications, etc. Here, our main objective is to estimate the impulse response from the response of the SSI to changes in sunspot area in order to better understand the nature of SSI variations.

\subsection{Impulse Response with Residual}
\label{subsect:res}
The impulse response model we have described is one that is commonly utilized in linear system theory \citep{antsaklis05}. However, the variability of solar irradiance may have different drivers, including some that are not directly caused by a single selected input $u(t)$. In other words, some part of the SSI variations may arise from processes independent of sunspots and products of their decay. To account for this, we extend the definition of the dynamical response by allowing two contributions:  1) the classical impulse response that relates output to input; and 2) an additional (but not necessarily small) residual contribution $r(t)$ that captures whatever is not described by the classical linear impulse response.  Note that we  do not impose any model for this residual, which merely describes what remains after modeling the convolution of a stimulus with the impulse response. Instead of equation~\ref{eq:impulse1}, we now have
\begin{equation}
y(t) = h(t)*u(t) + r(t),
\label{eq:impulse3}
\end{equation}
whose Fourier transform gives
\begin{equation}
Y(\omega) = H(\omega) U(\omega) + R(\omega).
\label{eq:impulse4}
\end{equation}
This system is fully characterized by $h(t)$ and $r(t)$. Here again, it is more convenient to work in Fourier space. In practice, for each frequency $\omega$, we have an ensemble of $M$ estimates of $\{U_k(\omega)\}$ and $\{Y_k(\omega)\}$, with $k=1, 2, \ldots, M$. The problem then amounts to finding for each frequency the two coefficients $\{H(\omega), R(\omega) \}$ that solve the over-determined set of equations $Y_k(\omega) = H(\omega) U_k(\omega) + R(\omega)$. This is typically done by total least squares, as classical least squares are not suitable because both $Y$ and $U$ have errors. The impulse response $h(t)$ and the residual $r(t)$  are then obtained by the inverse Fourier transform of the slope and the intercept of these lines, respectively. 

\begin{figure}[!htb] 
    \begin{center}
    \includegraphics[width=0.55\textwidth]{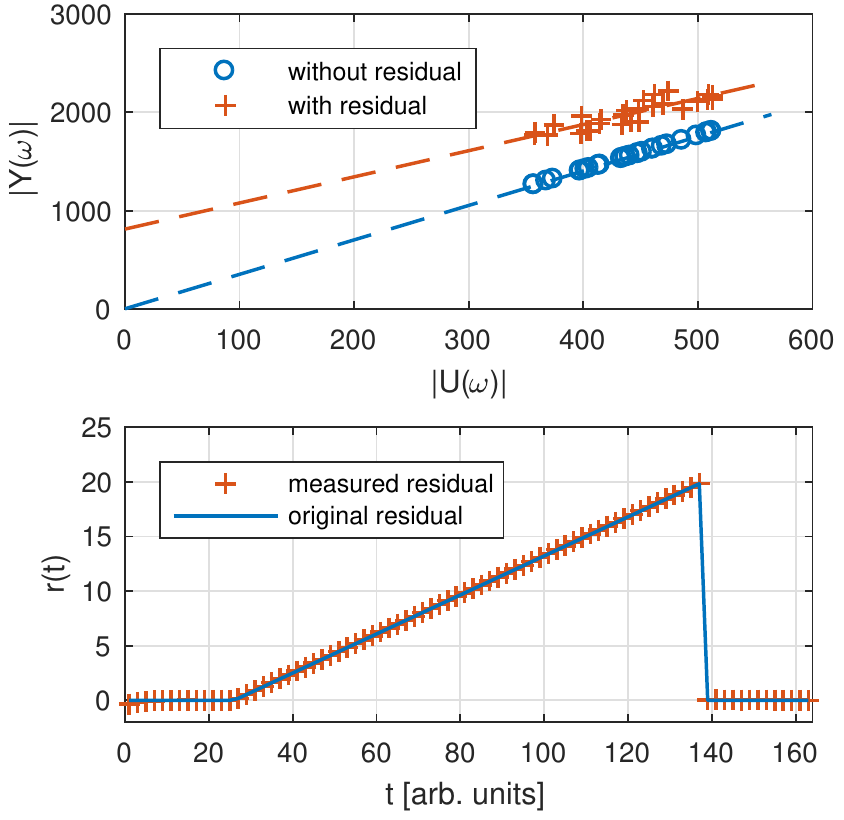}
    \end{center}
    
    \caption{Upper plot: Representation of $|Y(\omega)|$ versus $|U(\omega)|$ at a particular frequency for the example shown in Figure~\ref{fig_simul1} with and without an added artificial trend to the input time series. The regression is done in complex space; here we display the modulus for visualization purposes only. Also shown is the linear fit for each case, from which the impulse function and the residual can be estimated. Lower plot: The reconstructed residual (red crosses) closely matches the original trend (blue line) that was artificially added to the output. \label{fig_simul3}}
\end{figure}

To illustrate the importance of the residual term in equation~\ref{eq:impulse3}, we added an artificial trend to the output shown in Figure~\ref{fig_simul1} and estimated the resulting impulse response $h(t)$ together with the residual $r(t)$, again after adding noise to the in- and out-put. Figure~\ref{fig_simul3} compares the scatterplot of the output $|Y(\omega)|$ versus the input $|U(\omega)|$ at a given frequency with and without the added trend. When there is no trend in the data, the linear transformation intercepts the origin, as expected. However, this intercept is nonzero when the output has an additional trend. We thus have a simple criterion for both testing the presence of a residual and estimating it.

Figure~\ref{fig_simul3} also compares the reconstructed determination of the artificial trend to that actually applied, and confirms the ability of our methodology to properly extract both $h(t)$ and $r(t)$ from the observations, even when the latter are corrupted by noise. If no residual is allowed for in the model when there is one in the observations, then the impulse response will be biased and incorrect. In previous studies of the transfer function applied to solar irradiances, no such residual term was included, potentially biasing the determination of the impulse response by background residuals.  We find the inclusion of this residual term affects the resulting solar-response determinations and therefore we systematically include it in our model. 

System identification often relies on parametric approaches to estimate impulse responses because of their better noise immunity \citep{ljung97}. However, such approaches require the explicit design of a model of the response. While the impulse response can generally be described by a simple parametric model (e.g. an exponential in the example above), this is not true of the residual, whose variation in time can be arbitrarily complex. For that reason, parametric models are not appropriate for determinations of solar irradiance responses to various sources, as the measurements of each can include background effects that may be due to non-solar (i.e.~instrumental) causes. We thus utilize a more general non-parametric (Fourier) approach \citep[e.g.][]{schoukens09}.

\subsection{"Snapshots" Remove Solar-Rotational Variability}
\label{subsect:rot}
Aside from the inclusion of the residual term in equation~\ref{eq:impulse3}, there is a second important difference between our approach and that of PW. In the latter, all inputs and outputs used daily-cadence time-series data directly, thus incorporating both the desired solar-dynamical effects as well as unrelated solar-rotational effects (such as center-to-limb variations of contrasts of solar active regions as well as foreshortening).  These solar-rotational effects alter the true impulse-response function by modulating it with a 27-day oscillation, which, in addition, is wavelength-dependent. Although we could in principle deconvolve the two effects, it is considerably easier to reorganise the data in a way that eliminates the rotational modulation.  To do so, we consider a stroboscopic approach, taking a ``snapshot'' of the Sun once every solar rotation. Observing the Sun at this cadence allows us to concentrate on the dynamical evolution of active regions while strongly reducing the solar-rotational influences on their visibility; we thus separate most of the solar-dynamical response from the rotationally-induced variability. The tradeoff with this approach is coarser time resolution, since we cannot resolve timescales shorter than one solar rotation.

To illustrate this stroboscopic approach, consider a sequence of daily values $\{y(t_0), y(t_0+1), y(t_0+2), \ldots \}$. We extract from this a new record  $\{y(t_0), y(t_0+27), y(t_0+54), \ldots \}$ that is now sampled every 27 days.  One may also consider the sequence that is observed one day later:  $\{y(t_0+1), y(t_0+28), y(t_0+55), \ldots \}$ and, likewise, build additional records. In total, 27 different records can thus be generated. 

In this paper we analyze solar irradiance and solar proxy data from \textasciitilde12 years of observations (4401 daily values), giving  27 records of 163 time steps for each, with each time step corresponding to a different solar rotation.  These realizations are not fully independent; however, they contain exactly the same amount of information as the original time-series records. Let us therefore stress that we exploit the full data set, with no addition nor loss of information. 

Fortunately, because our 12 years of data correspond to nearly one solar cycle, each stroboscopic record is close to periodic.  This avoids the need to apply cumbersome windowing procedures that might otherwise be required before applying Fourier transforms.

The strength of the impulse-response approach we utilize lies in its ability to isolate and quantify the response of the Sun to changes in net disk-integrated sunspot area by observing the cumulative effect of an arbitrarily large number of spots for a sufficiently long time. Let us stress that we relate variations in the irradiance to variations in the sunspot area without distinguishing between short-lived and long-lived spots. In particular, a newly-emerging sunspot whose area exactly compensates a decaying sunspot will cause no variation in the irradiance. This is not exactly true, as the contribution to the SSI of features such as faculae is position-dependent. By using disc-integrated proxies we force the impulse response to describe an average response of the Sun. An important condition for this is the linearity of the solar response, which will be addressed in Section~\ref{sec_nonlinear}. The main consequence of the location-dependent effects of the sunspots combined with their varying locations of emergence on the solar disk will be a larger uncertainty in the values of the impulse response.

\section{Data}
\label{sec_data}

\subsection{The Solar Indices Used}

For solar irradiances, we utilize daily observations made by the Solar Radiation and Climate Experiment (SORCE) mission \citep{rottman05} from 115.5 to 2000 nm and by the Thermosphere Ionosphere Mesosphere Energetics and Dynamics (TIMED) mission \citep{woods05} from 29.5 to 115.5 nm. These two missions provide the longest time-interval for which the SSI was continuously monitored from the UV to the near-infrared, namely from 2003 May 15 to 2015 June 14. The spectral resolution is 1 nm from 29.5 nm to 310 nm and then progressively increases to 15.7 nm at longer wavelengths. For convenience, we concentrate on the irradiance integrated in six spectral bands: the extreme-ultraviolet (EUV) from 30-121 nm; the far-UV (FUV) from 122-200 nm; the middle-UV (MUV) from 200-300 nm; the near-UV (NUV) from 300-400 nm; the visible (VIS) from 400-600 nm; and the near-infrared (NIR) from 600-2000 nm. The SSI data contain a small fraction of observational gaps or outliers that are filled in by expectation-maximization \citep{ddw11}. This filling considerably facilitates the Fourier analysis and has no significant impact on the results.

The input to our solar irradiance impulse-response model should be a time series related to flux emergence of active regions on the visible side of the solar disk, as these regions are known to be causally related to the irradiance. In addition to the irradiances above, we consider five observables that provide inputs from which the irradiance is modeled: the daily sunspot-area (DSA) on the solar disk provided by the Royal Greenwich Observatory; the solar radio-flux at 10.7 cm (F10.7) from the Penticton Observatory; the solar radio-flux at 30 cm (F30) from the Nobeyama Observatory; the core-to-wing ratio (MgII) of the MgII line from the University of Bremen; and the total solar irradiance (TSI) from SORCE/TIM. The latter, being from a dedicated TSI instrument, is independent of the SSI provided by other instruments on the SORCE mission. Because the solar irradiance and these five observables are all disk-integrated measurements, the impulse function determined from each observable is the average response of the solar-radiative output without regard to the location(s) of the responsible active area(s) on the solar disk.

PW show that an ideal candidate for the input time series is the DSA, which is given as a percentage of the solar hemisphere. For convenience, we divide the official DSA by 100, so that its units are $10^{-4}$ of the fractional surface of the solar hemisphere (i.e.~100 $\mu$hem). In these units, a new sunspot typically  corresponds to an increase in $u(t)$ of the order of one unit; high levels of solar activity typically correspond to values of 10 and above.

For proper determination of the impulse response, it is important to consider the response of the Sun to a varying input and thus ignore any constant background or baseline. A natural choice for the latter is the level of activity corresponding to a quiet Sun.  For the SSI we define the response as $y(t) = \textrm{SSI}(\lambda,t) - \textrm{SSI}_{QS}(\lambda)$, where  $\textrm{SSI}_{QS}(\lambda)$ is the quiet Sun level that is estimated from periods having at least 30 consecutive days with no sunspots during the prolonged solar minimum of 2008-2009. For the DSA, the zero level is the natural baseline.

\subsection{The Need for Accommodating Residual Contributions}

\label{sec_resid}

\begin{figure}[!htb] 
    \begin{center}
    \includegraphics[width=0.55\textwidth]{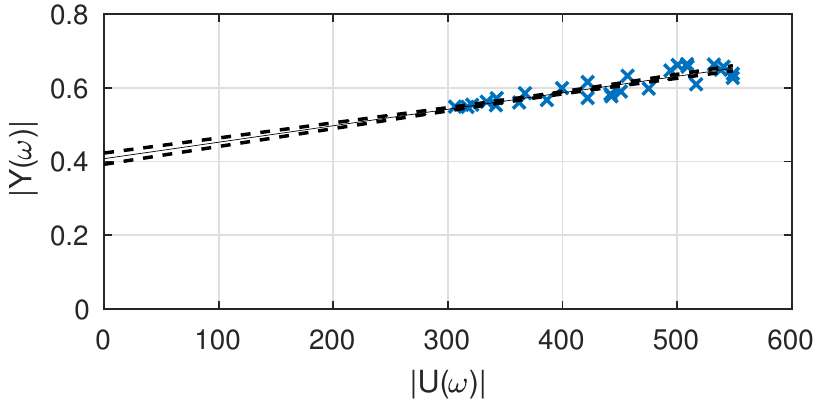}
    \end{center}
    
    \caption{Scatterplot of the lowest nonzero frequency (4401$^{-1}$ day$^{-1}$) of the Fourier component of the MgII index ($|Y(\omega)|$) versus that of the DSA ($|U(\omega)|$). Here we display the modulus solely for visualization purposes; the true regression is done in complex space. Also shown is the linear fit, from which the impulse function and the residual are estimated. The dashed lines correspond to $\pm$ one standard deviation of that fit. \label{fig_regression}}
\end{figure}

As shown in Figure~\ref{fig_simul3}, the scatterplot of $Y(\omega)$ versus $U(\omega)$ can clearly indicate whether a purely-linear impulse-response model is applicable and the need for the inclusion of a residual trend. Figure~\ref{fig_regression} similarly illustrates this using the DSA as input and the MgII index as output.  The nearly-linear relationship between the two quantities supports the validity of our linear and time-invariant impulse-response model. More importantly, it reveals the presence of a large offset $R(\omega)$ that cannot be neglected, which confirms the importance of including a residual term in our model. While Figure~\ref{fig_regression} illustrates the linear relationship for the lowest frequency only, qualitatively similar results are obtained at all other frequencies. Except for the lowermost frequencies however, the residual trend contains little power and the noise level is relatively high, which makes the linear relationship and in particular the offset less apparent.

\section{Results and Discussion}
\label{sec_discussion}

\subsection{Impulse Responses of Solar Indices to Sunspots}
\label{sec_discussion_impulse}

The central results of our study are the impulse responses $h(t)$ of the Sun to temporal changes in net sunspot area, which are representative of varying solar-surface magnetic activity. 

\begin{figure}[!htb] 
    \begin{center}
    \includegraphics[width=0.55\textwidth]{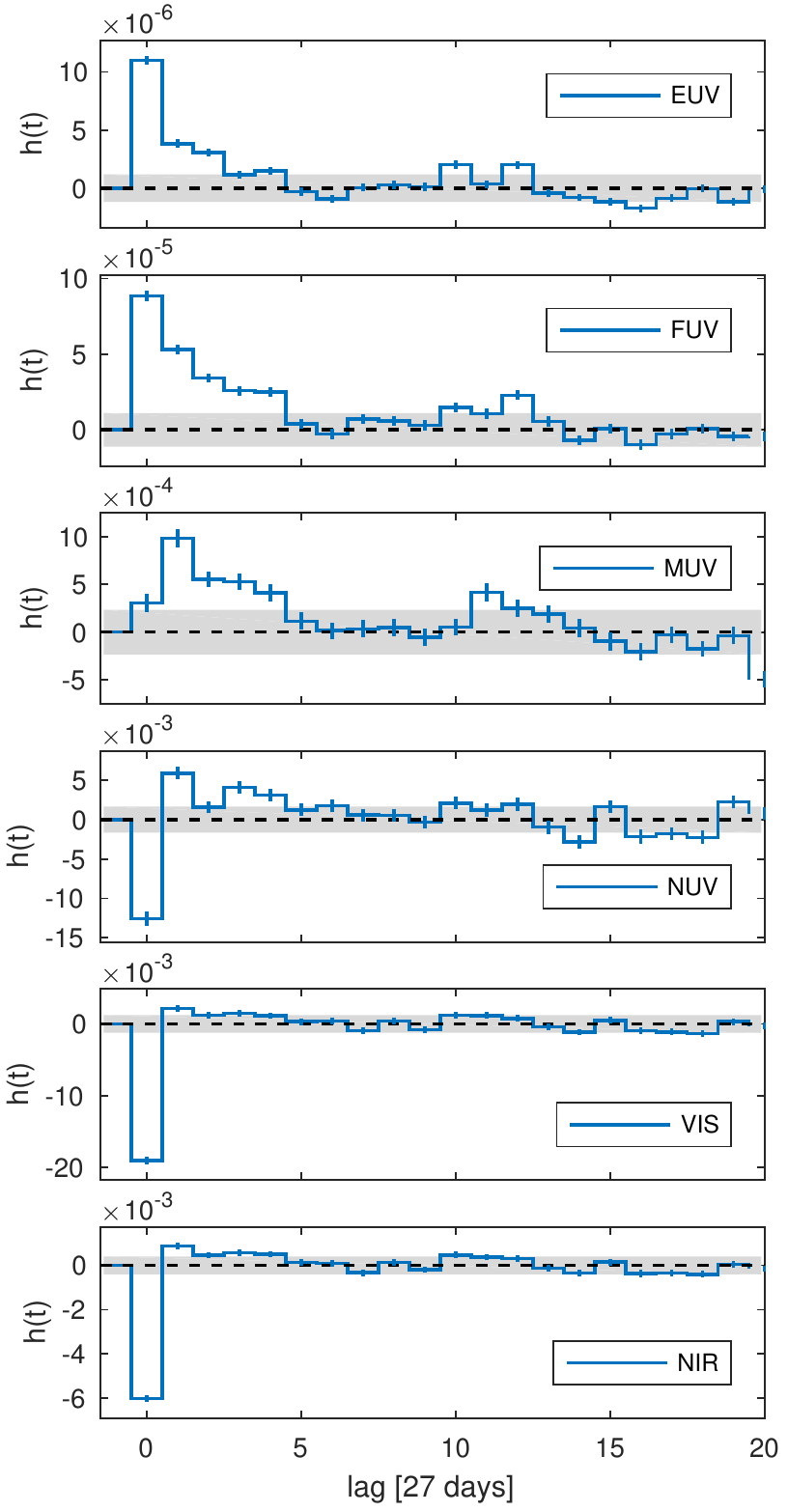}
    \end{center}
    
    \caption{The impulse response of the spectral solar irradiance due to the emergence of sunspots is plotted for six spectral bands. $h(t)$ is expressed in W~m$^{-2}$ per unit increase in the DSA, which is in 100 $\mu$hem. The shaded bands denote amplitudes for which the impulse response cannot be meaningfully distinguished from a randomly-varying input. The width of this band as well as the error bars plotted correspond to $\pm$ one standard deviation. \label{fig_impulse1}}
\end{figure}

\begin{figure}[!htb] 
    \begin{center}
    \includegraphics[width=0.55\textwidth]{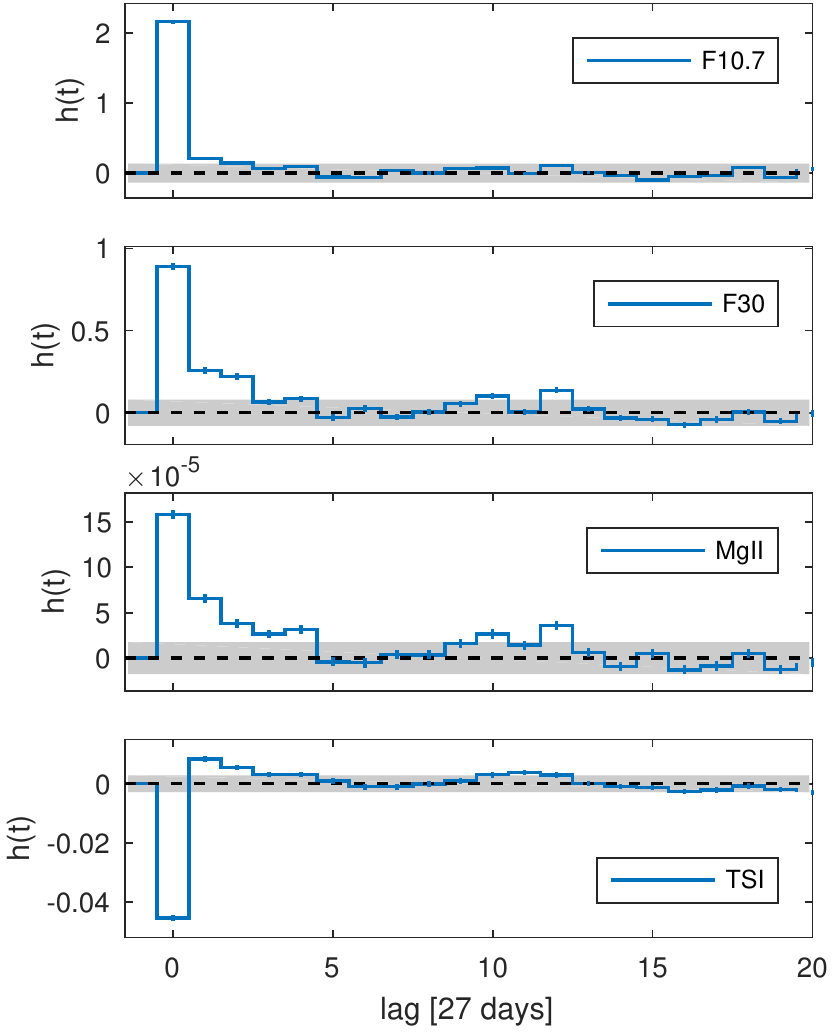}
    \end{center}
    
    \caption{The impulse responses due to sunspot emergence are plotted for various solar quantities: the radio flux at 10.7~cm (F10.7) and at 30~cm (F30); the MgII core-to-wing index; and the total solar irradiance (TSI). The shaded band and the errors again indicate $\pm$ one standard deviation. \label{fig_impulse2}}
\end{figure}

These impulse-response results are plotted in Figure~\ref{fig_impulse1} for the SSI over different spectral bands and in Figure~\ref{fig_impulse2} for the other solar observables. Each quantity's impulse response can be interpreted as the average variation in that quantity when the DSA increases by one unit (i.e.~100 $\mu$hem) for one solar rotation and then returns to its previous level for subsequent rotations. Although $h(t)$ is estimated for the full time-span (i.e.~163 solar rotations), we display it only for the first 20 solar rotations, of which only the first few are significantly nonzero.

Figure~\ref{fig_impulse1} illustrates that an increase in sunspot area on average initially generates an excess of emission in the shorter-wavelength UV bands and a deficit in the NUV and longer-wavelength bands, as indicated by the respectively positive (excess) and negative (deficit) values of $h(t)$ at $t=0$ in the impulse responses; Figure~\ref{fig_impulse2} shows a similar initial deficit for the TSI. The deficits are due to sunspot darkening initially exceeding the contributions from bright structures such as faculae and plage at these wavelengths, despite those bright structures generally appearing along with newly-emerging or increasing-size sunspots. We find these negative values occur only at $t=0$, meaning facular brightening overwhelms sunspot darkening within one solar-rotation period. 

Although this sunspot darkening is a known result \citep[e.g.][]{spruit00,froehlich04,foukal13}, our approach shows it in a more compact and quantitative way: $h(t)$ quantitatively describes the average change in measured SSI for a unit increase in sunspot area, and the method of using 27-day "snapshots" avoids the need for separately deconvolving solar-rotational effects. 

The impulse response for subsequent solar rotations ($t>0$) describes the decay of the solar response to flux emergence after the sunspot area has returned to its pre-emergent value. We find that all spectral bands in Figure~\ref{fig_impulse1} exhibit an excess of radiation after the initial $t=0$ response. This excess decays with an average time of $3.0 \pm 0.7$ solar rotations and is associated with the gradual conversion of remnants of the active region into enhanced magnetic network. 

The decay can be tracked at best for 3-4 rotations before the remnants of the active region are buried in the noise floor, which is represented in Figure~\ref{fig_impulse1} by a shaded band. This noise floor is obtained from surrogate data \citep{schreiber00}: we generate a large ensemble of synthetic records that have the same spectral properties and probability density as the original input and output but have randomized phases in Fourier space and thus are not causally related. From the standard deviation of samples of 600 impulse responses we obtain an indication of the level below which $h(t)$ cannot be meaningfully interpreted. With longer records and better statistics, this noise level shrinks and the decay can be tracked for longer periods; with the F10.7 index (for which observations started in 1947), for example, up to 7 rotations can be observed before reaching the noise floor.  

The number of nonzero terms in the impulse response is indicative of the number of solar rotations required for the signature in the SSI due to a changing sunspot area to vanish. Since only the 4--6 first terms of the impulse response are significantly nonzero, we conclude that the SSI has little or no memory of DSA variations by that same number of solar rotations. That is, the SSI cannot be delayed with respect to the DSA by more than a few months. We find no evidence for lags of several years or completely out-of-phase responses of the SSI, as suggested for example by \citep{woods15}.

The shape of the impulse response in Figure~\ref{fig_impulse2} helps clarify the behavior of solar proxies. The 30~cm radio flux (or F30 index), for example, has recently been advocated as a better proxy than the F10.7 for the UV \citep{ddw14} because it contains a stronger contribution from Bremsstrahlung, which is associated with UV-emitting bright loops above active regions. Figure~\ref{fig_impulse2} shows that both F10.7 and F30 have a bright peak at $t=0$ (which mainly comes from gyro-emissions that are associated with the active regions) followed by a decay that is considerably brighter in F30 than in F10.7. The comparison of the impulse responses shows that variability of the F30 index is indeed considerably closer to that of the MgII index than to the F10.7 index.

\subsection{The Residual Contribution}

The residual contribution $r(t)$ (see equation~\ref{eq:impulse2}) describes variations that cannot be reproduced by the time-invariant impulse response. Most of the spectral power of $r(t)$ is concentrated in low frequencies, primarily capturing slow variations on timescales of several months and longer. Note that the slow variation of the residual contribution is an observational result and is not imposed by our transfer function model. Figure~\ref{fig_simul3} demonstrates that $r(t)$ can describe sharp discontinuities as well.

\begin{figure}[!htb] 
    \begin{center}
    \includegraphics[width=0.65\textwidth]{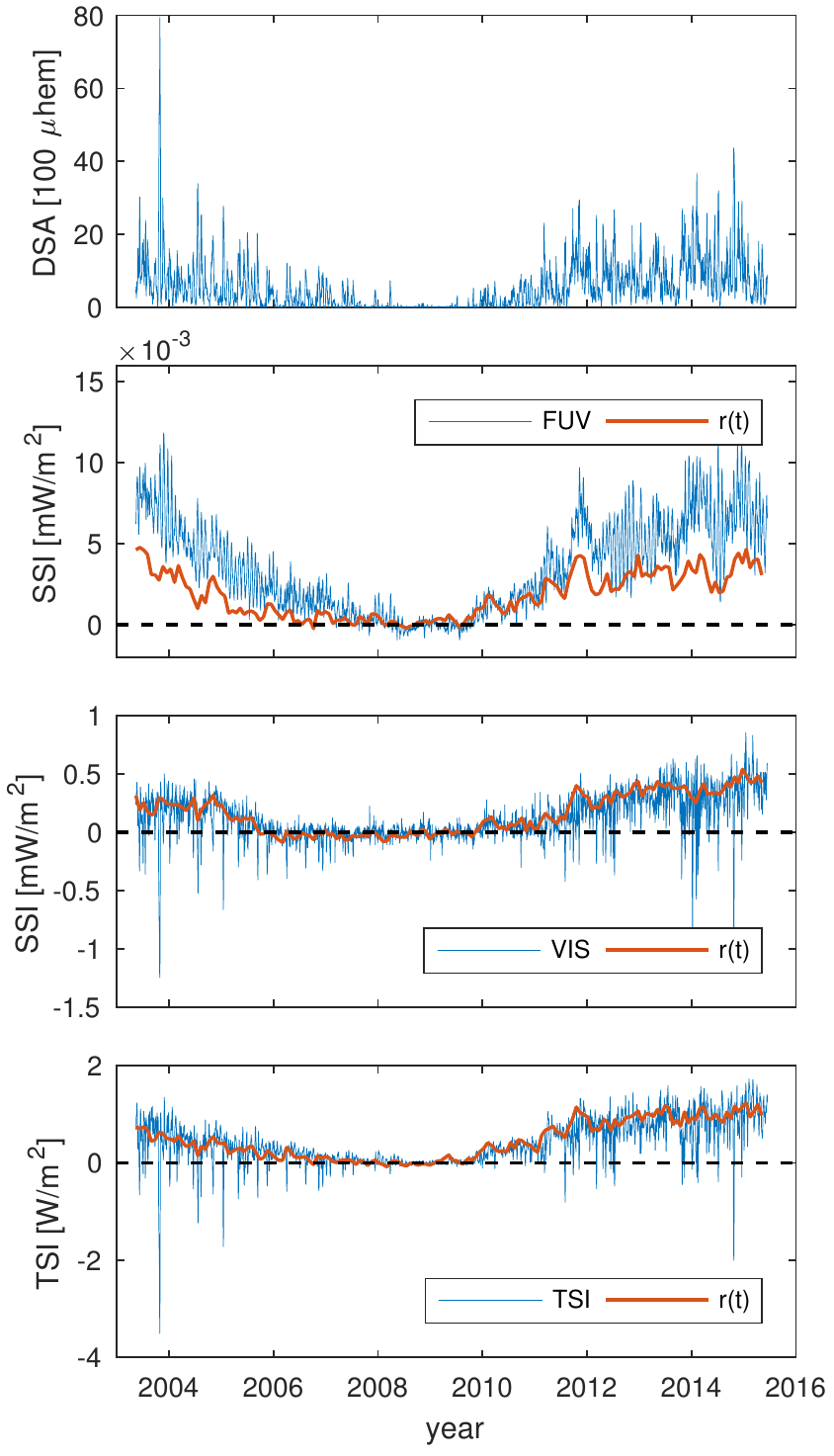}
    \end{center}
    
    \caption{The input DSA (upper plot) is modeled to three outputs, with the observations (blue) and their residual contributions $r(t)$ (red) shown: the solar irradiance in the FUV (second plot); the solar irradiance in the visible (third plot); and the TSI (fourth plot). The quiet Sun level has been subtracted from each quantity. The one standard-deviation uncertainty on $r(t)$ is approximately one eighth of its amplitude. \label{fig_residual}}
\end{figure}

Figure~\ref{fig_residual} illustrates the residual contribution $r(t)$ for two typical cases: in the FUV, where it captures approximately half of the long-term variability, and in the visible, where it accounts for almost all of the long-term variability. The relatively large amplitude of $r(t)$ is an important finding, since it means that a significant portion of the variability in the irradiance measurements cannot be reconstructed from changes in the sunspot area alone (at least, not with a linear model). This questions the validity of models that reconstruct the SSI from one single solar proxy, such as the net sunspot area or the even more-limited sunspot number; or it questions the long-term stability of the measurements themselves.

A nonzero residual contribution would be needed in the visible and near-infrared to explain any (controversial) out-of-phase variability in these bands with respect to solar proxies such as the DSA \citep{harder09} because our model relating the SSI to sunspot-area change does not produce lags in the SSI of more than a few months (as mentioned in Section~\ref{sec_discussion_impulse}). Thus such out-of-phase variability would instead be incorporated into the residual contribution and not the impulse response itself.

Let us stress again that in a non-parametric model, the residual contribution cannot be neglected as it was in previous studies since this affects the determination of the impulse response itself, as shown in Section~\ref{sec_resid}. The example shown in Figure~\ref{fig_regression} clearly indicates the need for using equation~\ref{eq:impulse3}. If, however, we ignore the residual contribution then reasonably good reconstructions can still be achieved by forcing equation~\ref{eq:impulse1} to the data, as in \citep{preminger07}. However, the quality of the reconstructed output in this case does not necessarily reflect the quality of the model parameters, with the appearance of long and unphysical tails in $h(t)$.

\begin{figure}[!htb] 
    \begin{center}
    \includegraphics[width=0.55\textwidth]{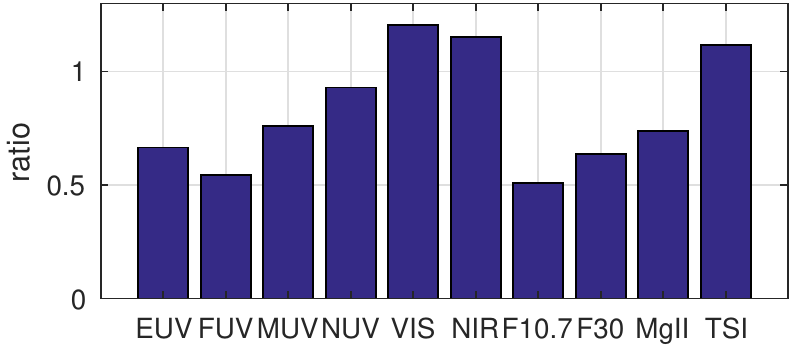}
    \end{center}
    
    \caption{Ratio between the amplitudes of $r(t)$ and $y(t)$ after decimating the latter to 27 days. Values larger than one are possible because we use here the standard deviation as a coarse measure of amplitude. \label{fig_ratio}}
\end{figure}

With a nonzero residual contribution such as we find, the SSI cannot be described solely by the sunspot areas because we have no model for describing how the residual contribution itself varies. Since numerous attempts have been made to infer past values of the TSI or SSI from sunspot data on a daily- to monthly-basis \citep[e.g.][]{solanki99,detoma01,pap02,balmaceda07,preminger07}, it is insightful to determine what spectral bands require a significant residual contribution. To quantify this, we compute the ratio between the amplitudes of $r(t)$ and of $y(t)$ after decimating the latter to the 27-day sampling rate of $r(t)$.  The results are summarized in Figure~\ref{fig_ratio}, which shows that the residual contribution represents at least 50\% of the long-term variation in the UV bands and approximately 100\% of the long-term variation in the visible and NIR.

The surprisingly large magnitude of the residual contribution may have several causes. As mentioned previously, this contribution is required to describe physical out-of-phase variations of the SSI on long timescales, namely those of several months and longer. 

Out-of-phase variations on shorter timescales have been well documented \citep[e.g.][]{vigouroux97,preminger05} and are captured by the impulse response itself. Longer-term variations may be caused by ephemeral regions, whose contributions vary cyclically with the solar cycle but start before and end after the corresponding cycle \citep{harvey92}. Geometrical effects may also contribute: since the average latitudinal position of sunspots decreases during each solar cycle, the response of the SSI to sunspot variations slightly differs between the rising and decaying phase of the solar cycle. This essentially means that the impulse response is not completely time-invariant. 

A different cause may be the presence of uncorrected instrumental drifts in the SSI, which the model cannot fit and thus defers to the residual contribution. Such drifts are more likely to occur in the visible and near-infrared bands, where the relative solar-cycle variability is lower than in the more energetic shorter wavelengths. However, since we find the residual is comparable in the visible band and the TSI, which share similar properties but have different long-term uncertainties (with the TSI measurements being considerably more accurate and stable on solar-cycle timescales), we conclude that instrumental effects are likely not a major contributor to the residuals shown.

Finally, since the residual contribution captures what the linear impulse response model cannot fit, what we observe may be the consequence of the limitations of this linear model with respect to a Sun that is not fully linear, time-invariant, and causal. These possible effects are discussed in more detail in Section~\ref{sec_nonlinear}.

We conclude here that the residual contribution, which represents an important part of the variability, probably captures both instrumental effects and physical processes that cannot be described by the impulse response. Therefore it is risky (if not impossible) to draw conclusions about long-term (i.e.~solar cycle) variability of the SSI and TSI simply by considering the linear and time-invariant short-term response to changes in sunspot area, because a significant portion of the solar variability, particularly in some wavelength ranges, is only partly related to these changes.

\textbf{Note added after publication :} We since found out that the large magnitude of the residual contribution may be partly caused by the violation of the Nyquist-Shannon-Kotelnikov sampling theorem. Indeed, by sampling the observations every 27 days, the energy contained in high-frequency variations pollutes the estimation of the transfer function, which affects the residual contribution. Preliminary tests show that this effect amplifies the residual contribution, but is not sufficient to explain its existence.


\subsection{Energy Balance Due to Sunspot-Area Variations}

\begin{figure*}[!htb] 
    \begin{center}
    \includegraphics[width=0.9\textwidth]{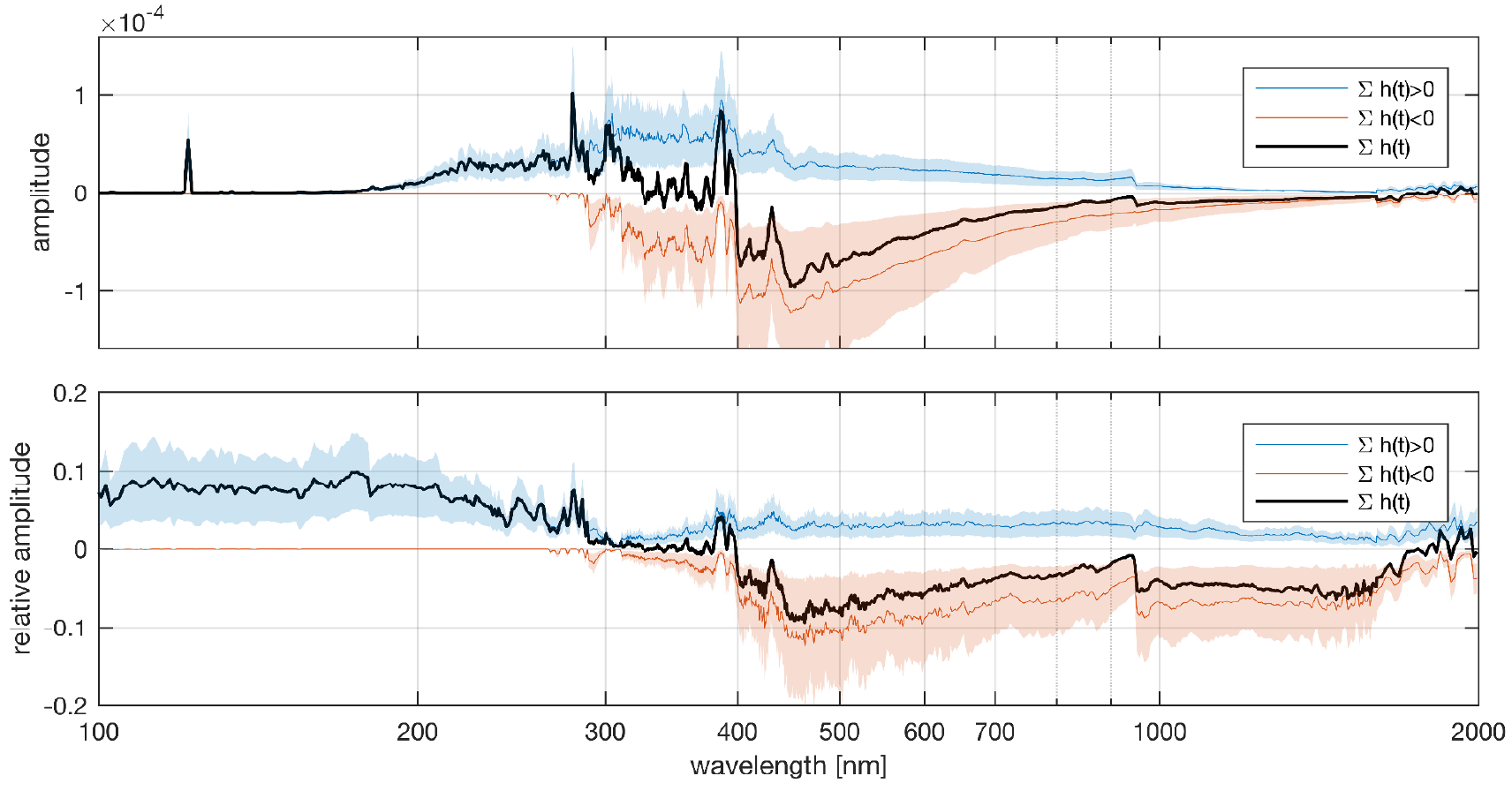}
    \end{center}
    
    \caption{Amount of spectral solar irradiance change per unit increase in the DSA. The blue (resp. red) curves indicate the amount of irradiance change that is caused by an excess (resp. deficit) of irradiance. The black curve is the sum of the two and expresses the total irradiance budget as a function of wavelength. The upper plot is in absolute units (W~nm$^{-1}$~m$^{-2}$ per 100 {$\mu$}hem) while in the bottom plot all quantities have been normalized with respect to the solar-cycle variability of the SSI. Uncertainties of $\pm$ one standard deviation are represented by shaded areas. For greater clarity, uncertainties are not indicated for the net irradiance budget (black); however they are of the same order of magnitude as the ones given for the excess and deficit. Some of the discontinuities, such as the dip near 930 nm, are instrumental, for instance coinciding with transitions between different detectors.
\label{fig_balance}}
\end{figure*}

In principle, the temporal integration of the impulse response in Figure~\ref{fig_impulse1} could enable energy-budget estimates and determine whether the average net contributions from facular brightenings exceed those from sunspot dimmings over the duration of the response. This is clearly the case for the shorter (EUV, FUV, and MUV) wavelength bands in Figure~\ref{fig_impulse1} and for the solar indices other than the TSI in Figure~\ref{fig_impulse2}, as these impulse responses are always positive. At longer wavelengths and for the TSI, the net contribution to brightness can be negative when integrated over the few solar rotations where the magnitude of $h(t)$ exceeds the noise limit, suggestive of a net energy loss due to the emergence of new sunspot-area flux. Temporal integrations of the responses for positive and negative values of $h(t)$ are shown as a function of wavelength in Figure~\ref{fig_balance} along with the sum of the two.

As the noise limit prohibits lengthier integrations, drawing conclusions on the net radiative energy balance due to increases in sunspot area remains uncertain. Additionally, extending these short-term energy-balance results to solar-cycle timescales, as suggested by Woods et al. \citep{woods15}, is even less certain since many solar-cycle effects are absorbed into the relatively-large residual component $r(t)$ rather than the impulse response itself. We thus conclude that it is particularly difficult to draw solar-cycle energy-balance conclusions via this linear impulse-response approach. While the approach can determine short-term (few solar rotations) net-energy responses to flux emergence for different wavelength bands, we find that the method is unable to determine in- or out-of-phase responses for the SSI on solar-cycle timescales.

\subsection{Significance of Precursors}

Our transfer function is causal in the sense that it assumes the SSI cannot start to vary before a sunspot appears. However, small precursors may exist, for example with ephemeral regions that occur before the sunspot emerges \citep{martin79}. The impulse response obtained by Preminger and Walton \citep{preminger05} actually shows variations in $h(t)$ for negative times. However, since their impulse response includes solar-rotational variations as well as contamination from (unfitted) residuals, an ambiguity in start time may explain this effect: in their model, $t=0$ corresponds to the occurrence of the maximum increase in sunspot area, which is influenced by sunspot position relative to the center of the disk (when its area is largest). Thus, increases in the the SSI may be observed at least a few days earlier (i.e.~for $t<0$) when the sunspot becomes visible on the east limb. Such ambiguities are greatly reduced in our approach of using 27-day snapshots to avoid such potential solar-rotational effects.

To check for the presence of precursors we estimated the impulse response after artificially delaying the solar response by an integer number of solar rotations, typically one to six. Although the impulse response fluctuates before the actual emergence of the sunspot, we never found these variations to be significant at the 0.05 level. We thus conclude that there is no evidence for precursors in the records we utilize here.

\subsection{Limitations - Nonlinearity in the Impulse Response}
\label{sec_nonlinear}

A major assumption behind our analysis is the linearity  of the solar response to sunspot-area variations.  While several studies suggest that this response should be mostly linear \citep{foukal90,lean95,preminger05}, nonlinear effects are expected to become significant for high levels of solar activity \citep{brown80,foukal98,solanki99}.

From physical considerations, the linear proportionality between DSA and SSI is expected to level off at high solar activity because the emergence of new sunspots is increasingly hindered by presence of nearby ones. To check for such a nonlinear signature in the impulse response we estimated the latter separately for three regimes of the solar cycle: weak activity (81-day average of the DSA  < 250 $\mu$hem), medium activity (250 <  DSA  < 750 $\mu$hem), and high activity (DSA > 750 $\mu$hem). The uncertainties on the results are larger than before because the sample sizes are smaller and the samples are not continuous in time.

\begin{figure}[!htb] 
    \begin{center}
    \includegraphics[width=0.45\textwidth]{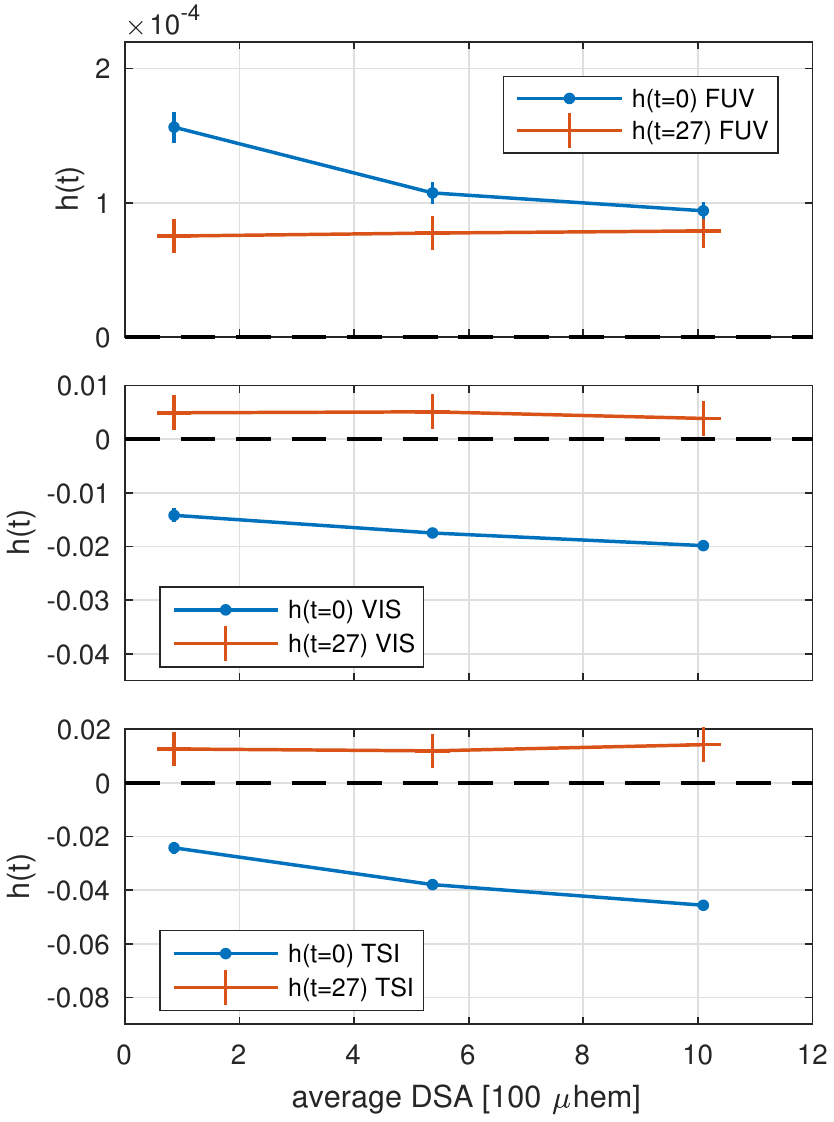}
    \end{center}
    \caption{Magnitude of the impulse response $h(t)$ at $t=0$ and one solar rotation later for the FUV band (upper plot), the visible band (middle plot), and the TSI (lower plot). Three values are shown, corresponding to three levels of solar activity. Error bars represent $\pm$ one standard deviation. \label{fig_nonlinear}
}
\end{figure}

Figure~\ref{fig_nonlinear} summarizes the impact of the average level of solar activity by showing the impulse response at two solar-rotation times, namely at $h(t=0)$ and at $h(t=27 \textrm{days})$. For both of these times, we plot the impulse response for the FUV, the visible, and the TSI for each of the three different net sunspot-area values, which are indicative of different levels of solar activity through the solar cycle. This plot reveals a nonlinear response for different phases of the solar cycle. 

For the UV bands, we find that the initial $t=0$ value of the impulse response, which is positive due to net solar brightening at these wavelengths, decreases by up to 40\% for the larger net sunspot areas typical of solar maximum. In the visible bands and for the TSI, where this initial response is negative due to overall solar darkening, the impulse response becomes stronger (that is, more negative) towards solar maximum. This means that an equal increase in sunspot area produces a stronger darkening at solar maximum than at solar minimum. Such a behavior of the impulse response function obtained for different activity levels indicates that the ratio of facular area to sunspot area is lower during periods of higher solar-activity levels. This is in line with previous observations, see for example Figure~2 in \citep{foukal93} and Figure~1 in \citep{foukal98}. 

Interestingly, after one or more solar rotations the impulse response is independent of the level of activity. This is because while the initial response is caused by an increase in sunspot area and any associated faculae, the later response is mostly caused by the decay products of the emerged sunspots, and those products do not depend on the level of solar activity.

Figure~\ref{fig_nonlinear} shows that the nonlinear response at $t=0$ is not just a matter of applying a static nonlinear correction to the input (e.g. using the square root of the DSA as input). If this had been the case, then the magnitude only and not the shape of the impulse response $h(t)$ would have depended on the level of solar activity. 

Figure~\ref{fig_nonlinear} indicates that the dynamics itself (i.e. the characteristic response time) is solar-cycle dependent. The properties of such nonlinear systems cannot be properly inferred from the usual scatter plots between the driver and the solar response; they require a  system identification approach with nonlinear models \citep[e.g.][]{nelles01}. Unfortunately such techniques are much more demanding in terms of data quality and volume; whether they can be meaningfully applied to the 12 years of data used here remains to be investigated.

Nonlinear effects may also manifest themselves in other aspects, such as a solar-cycle dependence of the characteristic response time, i.e.~an absence of time-invariance. Currently, with the relatively limited amount of available observations and the omnipresence of noise, it is unlikely that such second-order effects can be properly quantified. In this paper, while we acknowledge the possibility of these nonlinear effects, we thus consider only the average linear impulse response and deviations therefrom.


\section{Conclusions}
\label{sec_conclusions}

The prime motivation of this study was to use the powerful framework of linear system theory to better understand the response of the spectral solar irradiance (SSI) to sunspot-area variations.

We have discovered and presented several significant methodological improvements to the linear impulse-response model applied to sunspot-area changes (flux emergence) and the corresponding effects on the SSI. We have obtained a compact representation of the impulse response of the SSI to sunspot area that is independent of artifacts due to solar rotation via analyzing solar-rotational "snapshots." We have shown that a non-parametric model of the response of SSI cannot be entirely described by $h(t)$, but that a residual term is required. This important term, absent in prior such works, describes a major fraction of long-term variability. The existence of this residual term implies that one cannot describe the full variability of the SSI by solely using the sunspot area as input. This has implications on the use of empirical models that rely purely on the sunspot number: while such models perform well for short-term variability (days to months), they may fail to properly describe long-term (solar-cycle) variability.

Additionally, we have identified several physical results regarding the solar response to sunspot-area increases. The impulse response presented quantifies the global brightening observed in the UV. In the visible and near-infrared bands, an initial deficit of radiation is followed by a brightening, which can be observed for three to five solar rotations. Uncertainties associated with the relatively short (12 year) SSI-dataset time-span prevent observing potentially-lengthier decay responses. 

Nevertheless, we estimate a net energy budget for short-term variability as a function of wavelength, showing in which spectral bands sunspot darkening outweighs facular brightening. This energy budget includes uncorrected instrumental effects inherent in the data, which may limit its interpretation. 

We find the response of the SSI is nonlinear with the level of solar activity. This effect is mostly observed during the initial occurrence of the spot; the facular brightening that follows at solar-rotation times $t>0$ is independent of the level of activity. This indicates that the dynamical response of the Sun is nonlinear and cannot be corrected simply by applying a rescaling to the sunspot area.

There are two fundamental limitations to the conclusions presented here. {\it Data-length limitations} prevent us from drawing conclusions regarding solar-cycle variability since we have only relatively short durations of measurements of SSI variations. {\it Methodological limitations} of our linear model, which captures most of the short-timescale dynamics of the Sun, prevent it from reproducing variations equally well during solar maximum and solar minimum. A non-linear impulse-response model is required since the response of the SSI to sunspot area is not instantaneous, and therefore requires a convolutive model. This will be the subject of future endeavors based on the results presented here.


\subsection*{Acknowledgments}
We gratefully thank the instrumental teams for making their data available: the SORCE and TIMED teams,  Penticton Observatory, Nobeyama Radio Observatory (The Nobeyama Radio Polarimeters are operated by Nobeyama Radio Observatory, a branch of National Astronomical Observatory of Japan),  Royal Greenwich Observatory, and the University of Bremen. TD and MK thank CNES for supporting this study. GK appreciates funding from the NASA SORCE mission for his efforts. AS and VW received funding from the European Research Council under the European Union Horizon 2020 research and innovation programme (grant agreement No. 715947).


\end{document}